\documentclass[aps,prb,preprint,superscriptaddress,groupeaddress]{revtex4}
\usepackage{graphicx}
\usepackage{amsmath}
\usepackage{mathrsfs}
\usepackage{accents}
\begin{document}

\title{Superconductivity in an Al-Twisted Bilayer Graphene-Al Junction device}

    \author {Dingran Rui}
    \affiliation{Bejing Key Laboratory of Quantum Devices, Key Laboratory for the Physics and Chemistry of Nanodevices and Department of Electronics, Peking University, Beijing 100871, China}
    \author {Luzhao Sun}
    \affiliation{Center for Nanochemistry, Beijing Science and Engineering Center for Nanocarbons, Beijing National Laboratory for Molecular Sciences, College of Chemistry and Molecular Engineering, Peking University, Beijing 100871, P. R. China}
    \affiliation{Academy for Advanced Interdisciplinary Studies, Peking University, Beijing 100871, China}
    \author {N. Kang}
    \affiliation{Bejing Key Laboratory of Quantum Devices, Key Laboratory for the Physics and Chemistry of Nanodevices and Department of Electronics, Peking University, Beijing 100871, China}
    \author {Hailin Peng}
    \affiliation{Center for Nanochemistry, Beijing Science and Engineering Center for Nanocarbons, Beijing National Laboratory for Molecular Sciences, College of Chemistry and
    Molecular Engineering, Peking University, Beijing 100871, P. R. China}
    \author {Zhongfan Liu}
    \affiliation{Center for Nanochemistry, Beijing Science and Engineering Center for Nanocarbons, Beijing National Laboratory for Molecular Sciences, College of Chemistry and
    Molecular Engineering, Peking University, Beijing 100871, P. R. China}
    \author {H. Q. Xu}
    \email[Corresponding author: ]{hqxu@pku.edu.cn}
    \affiliation{Bejing Key Laboratory of Quantum Devices, Key Laboratory for the Physics and Chemistry of Nanodevices and Department of Electronics, Peking University, Beijing 100871, China}
		\affiliation{Beijing Academy of Quantum Information Sciences, Beijing 100193, China}
    \date{\today}

\begin{abstract}
We report on realization and quantum transport study of a twisted bilayer graphene (tBLG) Josephson junction device. High-quality tBLG employed in the device fabrication is obtained via chemical vapour deposition and the device is fabricated by contacting a piece of tBLG by two closely spaced Al electrodes in an Al-tBLG-Al Josephson junction configuration. Low-temperature transport measurements show that below the critical temperature of the Al electrodes ($T_c\approx1.1$ K), the device exhibits sizable supercurrents at zero magnetic field, arising from the superconducting proximity effect with high contact transparency in the device. In the measurements of the critical supercurrent as a function of perpendicularly applied magnetic field, a standard Fraunhofer-like pattern of oscillations is observed, indicating a uniform supercurrent distribution inside the junction. Multiple Andreev reflection characteristics are also observed in the spectroscopy measurements of the device, and their magnetic field and temperature dependencies are found to be well described by the Bardeen$-$Cooper$-$Schrieffer theory.
\end{abstract}

\maketitle
 Since the discovery of graphene\cite{Novoselov2004}, the unique properties, such as the presence of Dirac quasiparticles and linear energy dispersion in monolayer graphene\cite{Novoselov2005,Zhang2005}, have encouraged a vast of investigations of the materials. Superconductor-normal conductor-superconductor (SNS) Josephson junctions with the normal conductor made from graphene materials can be used to study the combined physics of the graphene materials and superconductors. The investigation of phase-coherent transport, such as nondissipative supercurrent (i.e., superconducting proximity effect), in such an SNS Josephson junction  is essential to understand the physics in graphene superconducting quantum devices. The advanced nanofabrication technology and the new emerging materials have been employed to realize various graphene Josephson junction devices\cite{Heersche2007,Du2008,Mizuno2012,Komatsu2012,Amet2016,Shalom2016,Lee2017,Zhu2017}. In particular, graphene is an excellent material system for building an SNS junction with a highly transparent normal conductor-superconductor interface, due to the stability of graphene without surface oxidation and the technology simplicity in contacting it by a superconductor. New phenomena arise when inducing the superconducting proximity effect into the graphene junction, such as the specular Andreev reflections near the Dirac point\cite{Beenakker2006,Efetov2016} and the interplay between the quantum Hall effect and the proximity induced superconductivity\cite{Amet2016}.

 \par Recently, twisted bilayer graphene (tBLG), formed as Bernal bilayer graphene with a certain rotating angle between upper and bottom layers, has attracted much attention\cite{Li2010,Yan2012,Cao2018,CaoYuan2018,Yankowitz2019}. This new emerging bilayer material is intriguing when considering the new degree of freedom of twisted angle. Theoretical calculations indicate that at a sufficiently large twisted angle, tBLG exhibits linear dispersion relation, which is similar to monolayer graphene. While at a small twisted angle, there can exist flat bands and charge localization in tBLG\cite{Bistritzer2011,Rozhkov2016}.
Many new phenomena are discovered by changing twisted angle in tBLG, such as the van Hove singularities in the density of state\cite{Li2010,Yan2012}, the strong correlated state and superconductivity\cite{Cao2018,CaoYuan2018,Yankowitz2019}, and the network of topological channels\cite{San2013,Efimkin2018,Huang2018,Rickhaus2018}.  In spite of the intrinsic superconductivity in tBLG with the twisted angle approaching the ''magic angle'', one may also induce the superconductivity into tBLG through the superconducting proximity effect in a Josephson junction setup. Such a tBLG Josephson junction is of great interest for the study of interplay between topological states and the superconducting proximity effect.

 \par In this work, we report on realization and transport measurement study of a tBLG SNS Josephson junction device. This is for the first time a report on a systematic study of low-temperate transport properties of a tBLG-based Josephson junction device, although some preliminary results have been recently reported by us in a conference\cite{Rui2019}. The high-quality material of tBLG employed in the device fabrication was obtained by chemical vapor deposition (CVD) and the device was fabricated on a Si/SiO$_2$ substrate by contacting a piece of CVD-grown tBLG with Al electrodes using standard nanofabrication technique. Transport measurements show that below the critical temperature of the Al electrodes, the device possesses highly transparent Al-tBLG interfaces and exhibits large nondissipative supercurrents. The magnetic field dependence of the critical supercurrent shows a Fraunhofer-like interference pattern, indicating a uniform supercurrent distribution inside the
junction. Multiple Andreev reflections are also observed in the device, showing a phase-coherent charge transport in the tBLG junction region, and their magnetic field and temperature dependencies are found to be well described by the Bardeen$-$Cooper$-$Schrieffer (BCS) theory.

\noindent
\textbf{EXPERIMENTAL METHODS}

\par
The material employed in this work was tBLG grown on a Cu foil in a low pressure CVD system. To ensure the bilayer graphene growth, high partial pressure of H$_2$ is needed to guarantee the formation of H-terminated edges in the first layer during the growth process, which would assist the carbon species go through the first layer \cite{Zhang2014}. Therefore, the tBLG growth was proceeded under 1000 sccm of H$_2$ (partial pressure was about 900 Pa) and 0.8 sccm of CH$_4$ at 1020 $^{\circ}$C for 40 min. Note that the interlayer twisted angles of as-grown tBLGs cannot be precisely controlled, but can be estimated by measuring relative orientations of the sharp edges of the two layers (both the upper and bottom layers in the tBLG films were in  hexagonal shapes and have pretty sharp edges).
CVD-grown tBLG films were transferred onto a heavily doped Si/SiO$_2$ substrate (with 300 nm in SiO$_2$ thickness) from the Cu foil.  Thus, we can estimate the twisted angle in each tBLG film through the relative orientations of the edges of the two layers. The Raman spectroscopy and transmission electron microscopy (TEM) measurements were also applied to confirm the structural properties of the tBLG films. In the Raman spectra shown in Figure 1a, we can see an enhanced intensity of the G-band with respect to the 2D band, a feature of tBLG arising from matching of the laser energy with the energy of van Hove singularities in tBLG\cite{Yin2016,Havener2012,KimK2012}. The inset of Figure 1a shows  a high resolution TEM image of a CVD-grown tBLG sample, where a moir\'e pattern is clearly observed, which further confirms that the sample is a twisted bilayer film.

\par
For transport measurements, Al-tBLG-Al hybrid junction and tBLG Hall bar devices were fabricated using standard nanofabrication technique. First, selected tBLG films were patterned by electron beam lithography (EBL) and reactive ion etching. Then, Ti/Al (5 nm/90 nm in thickness) electrodes were fabricated on the selected patterned tBLG samples by a second step of EBL and metal deposition via electron-beam evaporation. Figure 1b shows a schematic view of an Al-tBLG-Al junction device. Figure 1c shows an SEM image of a fabricated  device, where the tBLG junction size, i.e., the distance between the two Al electrodes, is $\sim$170 nm, and  the inset is an optical image of the device. Here, we would like to note that two contact pads are made to each Al electrode. Thus, electrical measurements can be made in a four-point setup to eliminating the influences of the line resistances of the measurement system.

All electrical measurements were carried out in a $^3$He/$^4$He dilution refrigerator with a base temperature of 10 mK, far below the critical temperature of the Al electrodes. The magnetic field was applied perpendicular to the tBLG film. Before discussion of the results of measurements of the Al-tBLG-Al junction, we need to check the transport properties of the CVD-grown tBLG films.  For this, a fabricated tBLG Hall-bar device on the same substrate was measured. Figure 1d shows the results of the measurements of the Hall bar at magnetic field B=2 T. It is seen that a few quantum Hall plateaus were well developed at this relatively low magnetic field, which implies that our transferred CVD-grown tBLG films were of excellent quality. The resistance values of the quantum Hall plateaus in tBLG with a sufficiently large twisted angle follow the expression of $\frac{1}{R_{xy}}$=$\frac{8 e^2}{h}$(n+$\frac{1}{2}$) (n=0,1,2,...) \cite{Yamagishi2012}. This is different from the quantum Hall resistances of AB-stack bilayer graphene, which can be written as  $\frac{1}{R_{xy}}$=$\frac{4 e^2}{h}$n (n=1,2,3,...) (n=1,2,3,...)\cite{Novoselov2006}. In our measurement, as shown in Figure 1d, the quantum Hall resistances of the tBLG film ($\sim$12 degrees in twisted angle) at B=2 T show well-developed quantized plateaus at $\nu$=4 and $\nu$=12, corresponding to an eight-fold degeneracy of the first excited Landau Level of tBLG, which is a unique feature for tBLG with a sufficiently large twisted angle\cite{Choi2011}.

\noindent
\textbf{RESULTS AND DISCUSSION}
\par
Now, we will present and discuss the low-temperature transport measurements of Al-tBLG-Al junction devices. Several Al-tBLG-Al junction devices were fabricated. The data presented below were acquired for the device shown in Figure 1c and other devices showed similar transport properties. Figure 2a shows the voltage drop $V$ measured between the two Al electrodes of the device shown in Figure 1c as a function of source-drain bias current $I_{sd}$ at different temperatures. It is seen that at temperatures below the critical temperature ($T_c\sim$1.1 K) of the Al electrodes, the device showed the proximity-induced superconductivity, i.e., a dissipationless Josephson supercurrent before $|I_{sd}|$ exceeded a critical value. For example, at T=60 mK, when the absolute value of $I_{sd}$ was smaller than critical current $I_c^L$ (in the negative current bias case) or critical current $I_c^R$ (in the positive current bias case), where $I_c^L$ and $I_c^R$ are indicated by arrows in Figure 2a, the voltage measured was zero and the current flowed through the tBLG SNS junction device without a voltage drop. When $I_{sd}$ was beyond the critical value of $I_c^L$ or $I_c^R$,  non-zero voltage drop was observed and the device switched from the superconducting state to the normal resistive state. As the temperature increased, the device was seen to switch from the superconducting state to the normal resistive state at smaller $I_c^L$ or $I_c^R$ and no superconductivity appeared in the device when the temperature reached 1 K.

\par
Figure 2b shows the temperature evolutions of the extracted critical supercurrents $I_c^L$ and $I_c^R$ for the device. Clearly, both $I_c^L$ and $I_c^R$ became smaller as the temperature was increased from 60 mK and completely vanished at T=1 K. It can also be seen that $I_c^L$ and $I_c^R$ were nearly equal (especially, this is the case when the temperature was above 0.4 K). Considering the quality factor Q of our junction was smaller than 1, the device was most likely in an overdamped regime\cite{Taboryski1999}. When compared with the extracted critical supercurrent-temperature ($I_c$-T) characteristics to the Eilenberger theory and the Usadel theory\cite{Veldhorst2012}, the shape of our experimental $I_c$-T curves is closer to that of a ballistic junction device. The inset of Figure 2b shows the temperature dependencies of $I_cR_n$ products. Here a similar trend as the temperature evolution of the critical current is seen. The extracted values of the $I_cR_n$ products at low temperatures is strongly suppressed, compared with the theoretically predicted value of $I_cR_n=\pi\Delta$/e (where $2\Delta$ is the superconducting energy gap of the Al electrodes) for the clean short junction limit\cite{Beenakker1992}. This reduction has been commonly reported in SNS junctions and may be attributed to the difference between the intrinsic critical supercurrent and the measured premature switching current\cite{Heersche2007,Xiang2006}.

Figure 3a shows the voltage drop between the two Al electrodes in the tBLG SNS junction device as shown in Figure 1c as a function of applied source-drain current $I_{sd}$ at T=60 mK and at different perpendicularly applied magnetic fields $B$. Here it is seen that the critical supercurrent in the device decreased as the magnetic field $B$ was increased up to 0.5 mT. Figure 3b shows  a 2D color plot of the differential resistance $dV/dI_{sd}$ of the device as a function of $B$ and $I_{sd}$ at T=60 mK. Here the dark blue areas correspond to the cases where the device was in the superconducting state with the critical supercurrent found at each magnetic field from a sharp increase in the differential resistance (bright color point). It is seen that the magnetic field dependence of the critical supercurrent oscillated and showed a Fraunhofer interference pattern, which can be well fitted by the theoretical prediction\cite{Barone1982} (red  dashed curve) of
\begin{equation}
I_c(B)=I_c(0)|\frac{sin(\pi\Psi(B)/\Psi_0)}{\pi\Psi(B)/\Psi_0}|,
\end{equation}
where $\Psi(B)$=$A_{J\!J}$B is the flux penetrating the tBLG junction area $A_{J\!J}$ and $\Psi(0)$=h/2e is the quantum flux. This good agreement between our experiment and the theory indicates that the supercurrent density in the device was uniformly distributed in the tBLG junction area. From the Fraunhofer pattern of the critical current oscillations, we can extract the period of the magnetic field $\Delta$B=0.53 mT. Considering the dirty London penetration length (about 60 nm) in the Al electrodes, this period is consistent with the area of the tBLG junction in the device.

Apart from the supercurrent measurements, the spectroscopy measurements were also carried out. Figure 4a shows the measured differential conductance $dI/dV_{sd}$ of the device shown in Figure 1c as a function of applied source-drain voltage $V_{sd}$ at B=0.8 mT, T=60 mK and $V_b=15$ V. Here a series of conductance peaks are seen at both positive and negative applied source-drain voltages, see for example, those peaks marked by red vertical dashed lines on the positive $V_{sd}$ side in the figure. These peaks arise from coherent multiple Andreev reflection (MAR) processes\cite{Taboryski1999,Octavio1983}. We note that a sharp conductance peak was also observed at zero bias voltage in the measurements. This peak can be attributed the appearance of the supercurrent in the device. The voltage positions of the MAR peaks in an SNS junction device are located approximately at $V_{sd}$=$2\Delta/ne$, where n is an integer and $2\Delta$ is the energy gap of the superconductor electrodes in the device. In our experiment, the MAR peaks of orders n=1,2 and 3 were clearly observed, as marked by the red vertical dashed lines in Figure 4a. From the voltage positions of these MAR peaks, we can extract a value of $\Delta\sim 0.13$ meV for the Al electrodes in our device.

\par
Figure 4b shows a 2D color plot of the differential conductance of the device as a function of $V_{sd}$ and B, where the MAR conductance peaks of orders n=1, 2 and 3 appear as bright color lines (or bands). It is seen that with increasing B, the MAR conductance peaks moved towards lower values of $|V_{sd}|$. When B reached 12 mT, the critical field of the Al electrodes, no MAR peaks were observable in our device. In an SNS junction device, the observed magnetic field evolutions of the MAR conductance peak positions follow the prediction of the BCS theory of
\begin{equation}
\Delta (B)\approx\Delta(0) \sqrt{1-(B/B_c)^2}.
\end{equation}
In Figure 4b, the dashed lines mark the predictions of the BCS theory for the magnetic field evolutions of the MAR peak positions of orders n=1, 2 and 3. Here, an excellent agreement between our experiment and the BCS theory is found.

\par
Figure 4c shows a 2D color plot of the differential conductance of the device as a function of $V_{sd}$ and T, where the MAR conductance peaks of orders n=1, 2 and 3 appear again as bright color lines (or bands). These conductance peaks moved towards lower values of $V_{sd}$ with increasing T and vanished when T approached 1 K. The temperature evolutions of these MAR peak positions follows  the BCS theory, approximately, as
 \begin{equation}
 \Delta (T)\approx\Delta(0) \sqrt{cos[\frac{\pi}{2}(T/T_c)^2]}.
\end{equation}
In Figure 4c, the dashed lines mark the BCS theory predictions for the temperature evolutions of the first three MAR peaks (n=1, 2 and 3). Again, an excellent agreement between the experiment and the BCS theory is achieved.

\par Finally, we would like to discuss the transparency of the Al-tBLG interfaces  in the device. This transparency  can be deduced from the excess current $I_{exc}$ obtained from the linear extrapolation of the I-V curves at high bias voltages. The transmission coefficient can be written as\cite{Octavio1983,Beenakker1988}
 \begin{equation}
 T_r=\frac{1}{1+Z^2},
\end{equation}
where Z is the dimensionless scattering parameter, which is given by the expression of $eI_{exc}R_n/\Delta$, and $I_{exc}$ can be determined from the intercept at the current axis of the I-$V_{sd}$ curve in the spectroscopy measurements. Based on this analysis, we obtain the  transmission coefficient of $T_r$= 0.79 for our device, indicating that the tBLG-Al contacts were highly transparent in the device.

\noindent
\textbf{CONCLUSION}

In summary, we have fabricated an Al-tBLG-Al Josephson junction device and have performed low temperature transport measurement studies  of the device.  The measurements show that the device exhibits sizable proximity-induced supercurrents and the Al-tBLG interfaces are highly transparent in the device. The measured critical supercurrent as a function of perpendicularly applied magnetic field exhibits a Fraunhofer-like intereference pattern, indicating a uniform distribution of the supercurrent
density in the tBLG junction region. Signatures of multiple Andreev reflections have also been observed in the spectroscopy measurements of the device.  The magnetic field and temperature dependencies of the multiple Andreev reflection characteristics agree excellently with the predictions of the BCS theory. This work is the first report on a Josephson junction device made from tBLG and provides a solid experimental background for further developments of hybrid tBLG-superconductor devices for novel physics and application studies.
\newpage

\newpage
\noindent
\textbf{Author Information}
\noindent
\par
\textbf{Corresponding Authors}
\par *Email: hqxu@pku.edu.cn
\par

\textbf{Notes}
\par The authors declare no competing financial interest.

\noindent
\textbf{Acknowledgements}

We thank the Electron Microscopy Laboratory, School of Physics, Peking University for the help on High-Resolution TEM characterization. We acknowledge financial supports by the Ministry of Science and Technology of China through the National Key Research and Development Program
of China (Grant No. 2016YFA0300601 and 2017YFA0303304), the National Natural Science Foundation of China (Grant Nos. 11874071, 11774005, and 11974026), and the Beijing Academy of Quantum Information Sciences (Grant No. Y18G22). \\

\noindent{\bf References}


\clearpage

\begin{figure}
\begin{center}
\includegraphics[width=6.5in]{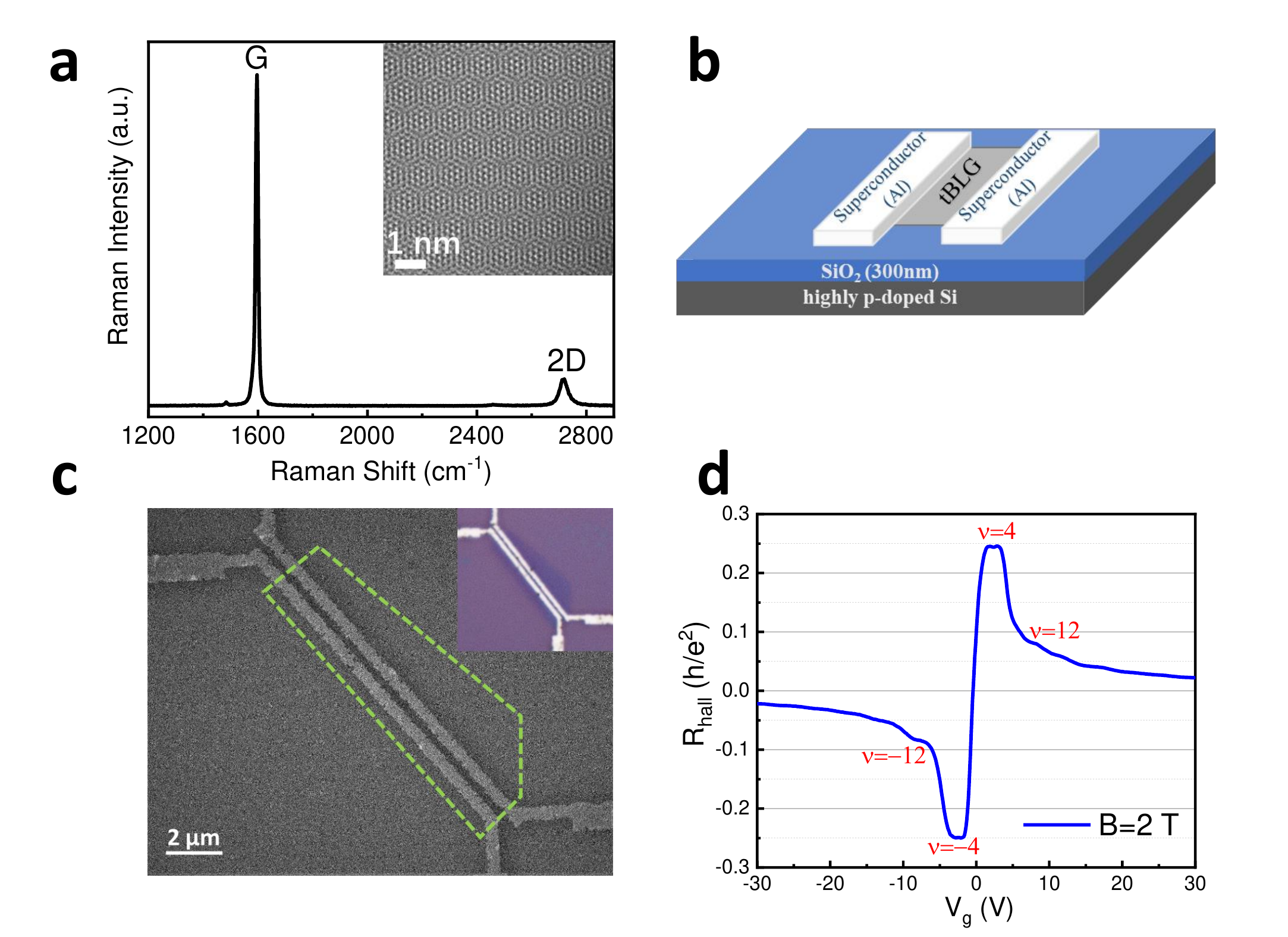}
\caption{
\textbf{a.} Raman spectra of a tBLG sample with a twisted angle of about 12 degree between the two graphene layers, taken under an incident laser with a wavelength of 514 nm. The inset displays a high-resolution TEM image of a tBLG sample with a twisted angle of about 12 degree. The scale bar in the inset corresponds to 1 nm.
\textbf{b.} Schematic view of a tBLG Josephson Junction device fabricated on a SiO$_2$/Si substrate, in which a CVD-grown tBLG film is contacted by two superconducting Al electrodes.
\textbf{c.} SEM image of the tBLG Josephson junction device studied in this work. Here the scale bar corresponds to 2 $\mu$m. The green dashed line shows the contour of the tBLG sample with a twisted angle of about 12 degree and the separation between the two Al contact electrodes is about 170 nm. The inset displays an optical image of the device.
\textbf{d.} Quantum Hall resistance of a CVD-grown tBLG film at a magnetic field of 2 T at T=1.9 K, indicating that our CVD-grown tBLG samples are of good quality.
}
 \label{fig1}
\end{center}
\end{figure}

\clearpage

\begin{figure}
\begin{center}
\includegraphics[width=6.5in]{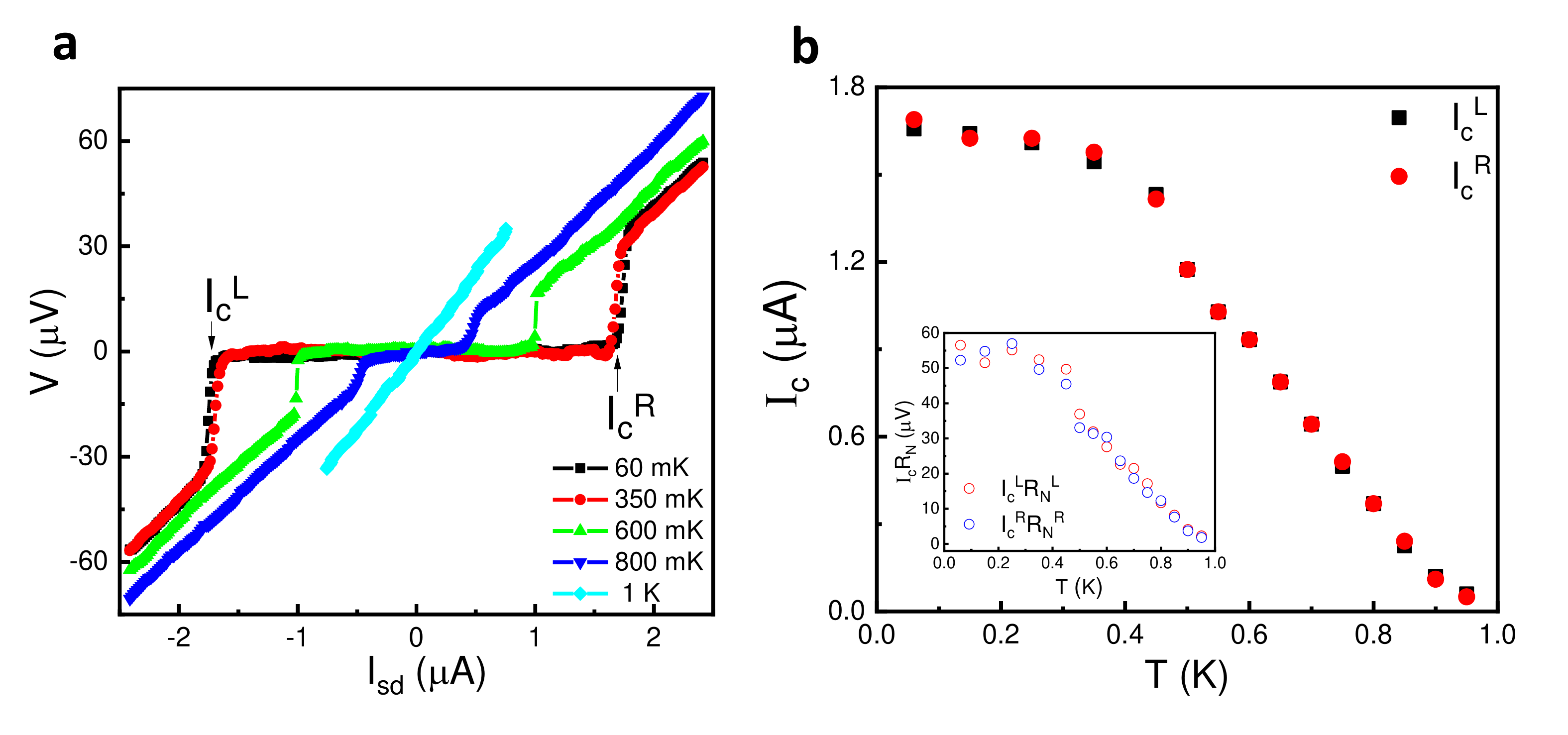}
\caption{
\textbf{a.} Measured voltage-current curves for the Al-tBLG-Al Josephson junction device as shown in Fig.~1\textbf{c} with back gate voltage set at $V_g=-3.7$ V at different temperatures. The critical currents $I_c^L$ and $I_c^R$ extracted from the measurements are indicated by the arrows in the figure.
\textbf{b.} Critical currents $I_c^L$ and $I_c^R$ plotted against temperature T for the device with the back gate voltage set at $V_g=-3.7$ V. The inset shows the corresponding $I_cR_n$ products as a function of temperature.
}
\label{fig2}
\end{center}
\end{figure}

\clearpage

\begin{figure}
\begin{center}
\includegraphics[width=6.5in]{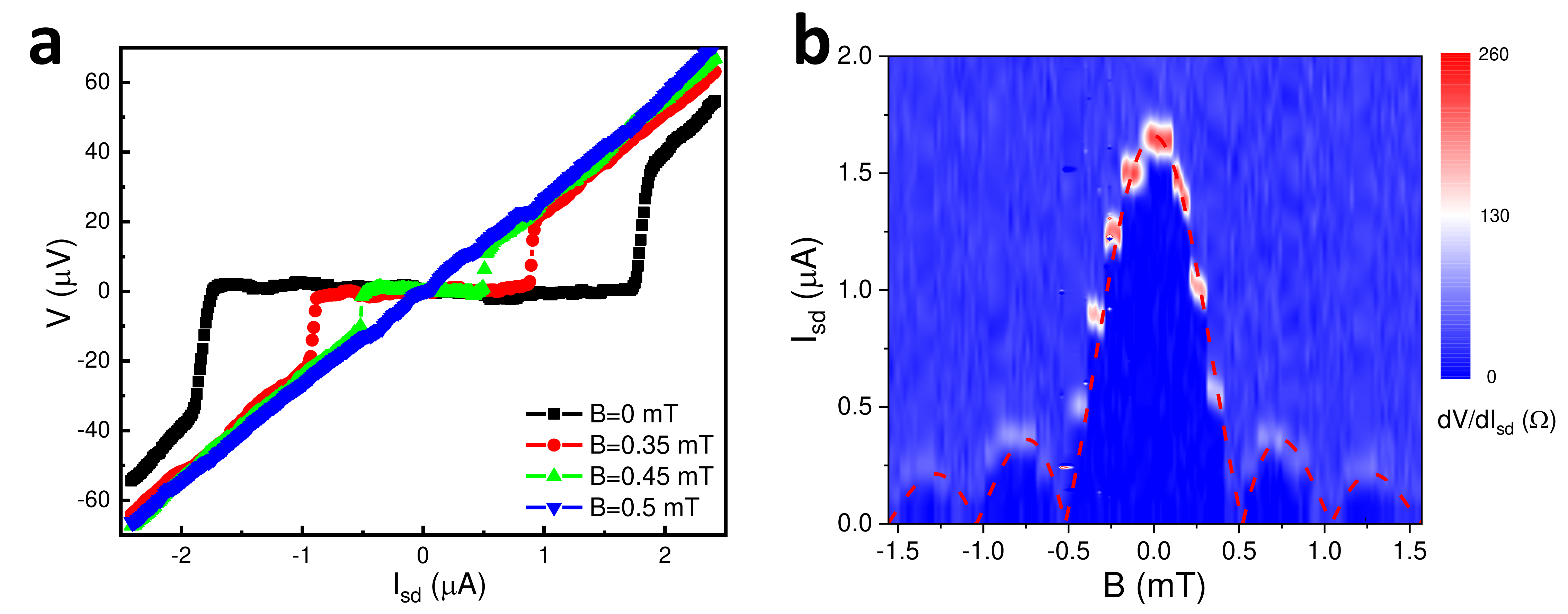}
\caption{
\textbf{a.} Measured voltage-current curves for the Al-tBLG-Al Josephson junction device with the back gate voltage set at $V_g=30$ V at $T=60$ mK and at different perpendicularly applied magnetic fields.
\textbf{b.} Differential resistance of the device measured as a function of the applied source-drain current and the magnetic field at $T=60$ mK and $V_g=30$ V. The dashed line corresponds to a calculated Fraunhofer pattern of the supercurrent in a corresponding ideal Josephson junction.
}
\label{fig3}
\end{center}
\end{figure}

\clearpage

\begin{figure}
\begin{center}
\includegraphics[width=6.8in]{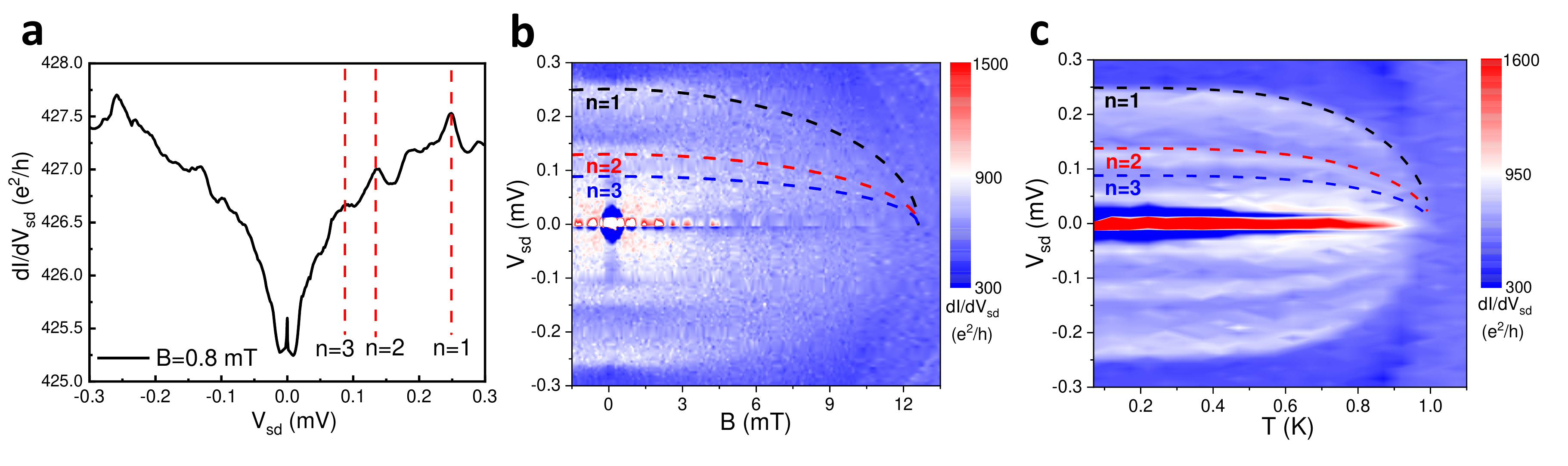}
\caption{
\textbf{a.} Differential conductance measured for the device as shown in Fig.1\textbf{c} as a function of applied source-drain voltage $V_{sd}$ at B=0.8 mT, T=60 mK and $V_g=15$ V. The red dashed lines show $V_{sd}$ values at which multiple Andreev reflection (MARs) peaks would appear. Here only the positions of $V_{s}=2\Delta/ne$  with n=1, 2, and 3 are shown.
\textbf{b.} Differential conductance measured for the device as a function of bias voltage $V_{sd}$ and magnetic field B at $T=60$ mK and $V_g=15$ V. The colored dashed curves are the predictions for the magnetic field evolution of the differential conductance peaks arising the MARs of orders n=1, 2 and 3.
\textbf{c.} Differential conductance measured for the device as a function of bias voltage $V_{sd}$ and temperature $T$ at $V_g=15$ V and zero magnetic field. The colored dashed curves are the predictions for the temperature evolutions of the differential conductance peaks arising from the MARs of orders n=1, 2 and 3.
}
\label{fig4}
\end{center}
\end{figure}

\clearpage

\end{document}